\documentclass[proceedings, preprint]{rmaa}
\usepackage{natbib}

\usepackage{paralist}

\usepackage{psfrag,color}

\SetYear{2009}
\SetConfTitle{Interferometry of hot stars}

\title{Infrared interferometry of massive young stellar objects} 

\author{
  W. J. de Wit,\altaffilmark{1} 
  M. G. Hoare,\altaffilmark{1}
  R. D. Oudmaijer,\altaffilmark{1}
  and T. Fujiyoshi\altaffilmark{2}
}
\altaffiltext{1}{School of Physics and Astronomy, University of Leeds,
Woodhouse Lane, Leeds LS2 9JT, UK (w.j.m.dewit@leeds.ac.uk).}

\altaffiltext{2}{Subaru Telescope, National Astronomical Observatory of Japan,
Hilo, Hawaii, USA}

\shortauthor{de Wit et al.}
\shorttitle{Interferometry of MYSOs}

\listofauthors{W. J. de Wit, M. G. Hoare, R. D. Oudmaijer \& T. Fujiyoshi}
\indexauthor{de Wit, W. J.}
\indexauthor{Hoare, M. G.}
\indexauthor{Oudmaijer, R. D.}
\indexauthor{T. Fujiyoshi}

\abstract{We discuss VLTI AMBER and MIDI interferometry in addition to single-dish
Subaru observations of  massive young stellar objects. The observations probe 
linear size scales between 10 to 1000\,AU for the average distance of our sources.}

\resumen{En este trabajo, presentamos observaciones interferometricas y de
plato unico de objetos estelares muy jovenes y masivos, que fueron realizadas
con los instrumentos AMBER y MIDI en el VLTI y con el telescopio
Subaru. Dichas observaciones ofrecen la posibilidad de explorar las regiones
centrales de estos objectos, cuyos tamanos lineales varian entre 10 y 1000 AU
para las distancias tipicas de estas fuentes.}

\addkeyword{Stars: formation}
\addkeyword{Stars: early type}
\addkeyword{Infrared: stars}
\addkeyword{ISM: clouds}
\addkeyword{Techniques: interferometric}

\begin{document}
% Typeset article header
\maketitle

\section{The MYSO environment}
\label{intro}
The massive YSO (MYSO) evolutionary phase is probably the earliest phase of a
nascent hot star in which the various circumstellar phenomena can be
identified and studied in detail at milli-arcsecond (mas) resolution in the
infrared (IR). MYSOs lack (compact) \ion{H}{II} regions from which one can
infer a high mass accretion rate. Either the star is extended and thus
relatively cool \citep{2009ApJ...691..823H}, or the
accretion quenches the \ion{H}{II} region \citep{1995RMxAC...1..137W}. It is not clear
which of these two scenarios applies. The ubiquitous presence of molecular
outflows \citep{2005IAUS..227..237S}, and in a few cases
collimated jets \citep[e.g.][]{2009ApJ...692..943C} all
point to on-going accretion in such systems.  Observations also indicate that
disk structures are prominent features during the MYSO phase \citep[e.g.][]{2004A&A...427L..13B}. This evidence suggests that the mass
accretion process proceeds (at least in parts) through a circumstellar disk,
which is substantiated by numerical models \citep[e.g.][]{2009Sci...323..754K}. 
 Disks are the likely source of an
ionised wind with modest escape velocities traced by near-IR
recombination lines \citep[e.g.][]{1995MNRAS.272..346B} and
high-resolution radio observations \citep{2006ApJ...649..856H}.
These defining components of the MYSO environment are all buried within a
thick protostellar envelope, and they suggest a picture of massive star formation
(SF) in many ways analogous to low-mass SF. We recall however that known MYSOs do not
correspond to the most massive ZAMS O-type stars observed, but probably to
early B or late O-type stars neglecting any multiplicity. Accreting stars with
$\rm M>25\,M_{\odot}$ remain elusive. Here we present high-angular resolution
observations of MYSOs in an attempt to understand their close environment on
scales between 10 and 1000\,AUs.

\section{At 1000 AU: the onset of rotation?}
The protostellar envelope plays a critical role in massive SF. 
Observations can reveal whether it is in collapse or in 
force balance.  Analysis of single-dish (sub)millimeter observations show
that the density distribution at 10,000\,AU is best described
by a radial powerlaw with powers between $-2.0$ to $-1.5$
\citep[e.g.][]{2002ApJS..143..469M}. This could correspond to 
the static outer envelope of an inside-out
gravitional collapse. In \citet{2009A&A...494..157D} we
present a survey of 14 well-known MYSOs at 24.5\,\micron~with the 8.2 meter
Subaru telescope probing size scales of $\sim 1000$\,AU. We find that 
the resolved envelopes follow a 
relatively shallow density powerlaw with exponent $-1$. Such a powerlaw is consistent with
the static outer part of an infalling logatropic sphere
\citep{1997ApJ...476..750M}, however the 1000\,AU size scale is too
small. Instead, we suggest that the observations probe the region where rotation becomes
dominant over infall and the transit occurs from a protostellar envelope to
a disk.

\section{At 100 AU: MIDI observations of W33A}
\begin{figure}[!t]\centering
  \includegraphics[angle=90,width=\columnwidth,height=0.75\columnwidth]{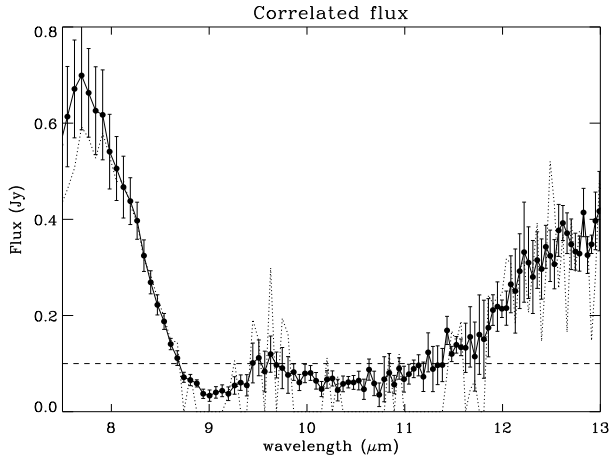}
  \caption{MIDI correlated flux spectrum of W33A. Flux levels below the dashed
  line must be considered upper limits (from de Wit et al. 2007).}
  \label{w33a}
\end{figure}

The VLTI-MIDI instrument is 
%a two-beam recombiner that operates in the N-band (8-13
%micron) and delivers spectrally dispersed visibilities and differential
%phases. The system is 
sensitive to 300\,K continuum emission, which in the
MYSO environment  could either be located in a circumstellar
disk, in the protostellar envelope or possibly in the outflow cavities
\citep{2005ApJS..156..179D}. In Figure\,\ref{w33a} we show the correlated MIDI flux
spectrum of the MYSO W33A ($\rm L=4\,10^{4}\,L_{\odot}$, $\rm 3.8\,kpc$). The
spectrum is dominated by the exceptionally strong silicate absorption feature,
indicating a very large dust optical depth. The corresponding visibilities
reveal that the emitting region increases linearly with wavelength from
120\,AU at 8\,\micron~to 240\,AU at 13\,\micron. This is consistent with
temperature falling off with distance. Simultaneous modelling of the
visibilities and observed spectral energy distribution with spherical dust
radiative transfer models leads us to conclude (de Wit et al. 2007) that the data is consistent
with a shallow dust density distribution (exponent $-0.5$ to $-1.0$). We also
find that a better fit is obtained if we decrease the effective temperature of
the central object to early A-type, making the underlying star effectively a
supergiant. The latter result is similar to the MIDI result
presented for the MYSO M8E-IR by \citet{2008ASPC..387..132L}.

\section{At 10 AU: AMBER observations of G310}
VLTI-AMBER is the three-beam recombiner operating in the $JHK$-bands delivering
spectrally dispersed interferometric observables. We have obtained fringes of the MYSO G310
($\rm L=2\,10^{4}\,L_{\odot}$, $\rm 3.3\,kpc$) in the low spectral resolution mode on 
the U2-U3 and U3-U4 baselines. The visibilities correspond to a ring size
of $\rm \sim 2.5\,mas$ or $\rm \sim 8\,AU$ at the adopted distance of G310.  This
size is consistent with that expected for a dust disk
truncated by a sublimation temperature of 1000\,K \citep[see][]{2002ApJ...579..694M}. The size of the emitting region is constant with
wavelength. At $K$-band scattering of the particles is relatively efficient, and a large
over-resolved component corresponding to the protostellar envelope is likely
to be present (the resolved emission blobs generally seen in single dish
images). However the
AMBER fringes show that imaging information on scales as small as 10\,AU are
within reach, probing the important region where the mass accretion is
actually taking place.

\vspace{-0.1cm}
\begin{center}
DISCUSSION
\end{center}
\vspace{-0.25cm}
\noindent
{\it H. Zinnecker}: Are there any MYSOs for which all three presented types of
data can be taken? \\
{\it W.J. de Wit}: We are working on that.\\
\noindent
{\it H. Zinnecker}: To what extend can you ignore the incidence of binaries in
your  fits to the 24.5\micron~data?\\
{\it W.J. de Wit}: In most cases the envelope is spherical to first
order. However we have clear examples of multiple sources within
a common envelope. \\
\noindent
{\it D. Setia Gunawan}: How significant is rotation in MYSO evolution, and how
does this relate to disk formation/binary formation?\\
{\it W.J. de Wit}: Rotating structures have shown to be present in massive SF
with molecular line observations. At what
point and in how far this determines the formation of a disk/binary is
not known.
%\bibliographystyle{rmaa}
%\bibliography{ihot}

\end{document}